\documentclass[12pt]{iopart}

\usepackage{graphicx}
\usepackage[colorlinks]{hyperref}
\usepackage{ulem}
\usepackage{color}

\begin{document}

\title{Vortex interactions in the collision of Bose-Einstein condensates}

\author{Tao Yang$^{1,2}$, Andrew J Henning$^{2,3}$ and Keith A Benedict$^2$}
\address{$^1$Institute of Modern Physics, Northwest University, Xi'an 710069, P. R. China}
\address{$^2$School of Physics and Astronomy, University of Nottingham, Nottingham NG7 2RD, United Kingdom}
\address{$^3$Present Address: National Physical Laboratory, Teddington, Middlesex, TW11 0LW, United Kingdom}
\ead{yangt@nwu.edu.cn}

\begin{abstract}
We investigate the effects of vortex interaction on the formation of interference patterns in a coherent pair of two-dimensional Bose condensed clouds of ultra-cold atoms traveling in opposite directions subject to a harmonic trapping potential. We identify linear and nonlinear regimes in the dipole oscillations of the condensates according to the balance of internal and centre-of-mass energies of the clouds. Simulations of the collision of two clouds each containing a vortex with  different winding number (charge) were carried out in these regimes in order to investigate the creation of varying interference patterns. The interaction between different vortex type can be clearly distinguished by those patterns.
\end{abstract}

%\begin{keyword}
%Vortex interaction \sep Bose-Einstein condensates \sep Interference
%\end{keyword}


\maketitle

\bigskip

\section{Introduction}

The collision and merging of Bose-Einstein condensates (BECs) using ground state or split clouds has been widely studied in atomic interference \cite{Science.275.637,PRL.66.2693,PRA.64.063607,PRL.92.050405,PRA.68.013611,PRA.81.043608,PRA.87.023603,PRA.88.043602,PRL.95.190403,PRL.100.100402}, in order to carry out measurements via ultraprecise interferometry \cite{PRL.78.2046,PRL.89.140401,PRL.94.090405}. The interference between condensates also provides a useful tool in order to create and detect topological states with vorticity. The quantization of circulation of the fluid, leads to quantized vortex states which affect the interference properties. Several ways of creating vortices experimentally in a BEC have been suggested, such as quantum phase manipulation \cite{PRL.83.2498,PRL.89.190403}, rotating traps \cite{PRL.84.806,Science.292.476,PRL.87.210403}, turbulence \cite{PRL.87.080402} and dynamical instabilities \cite{PRL.86.2926,Science.293.663}.The first two of these methods are analogous to techniques in more traditional superfluids such as liquid $^4$He.

The interference of two condensate clouds under expansion, one in a vortex state and one in the ground state, has been studied to develop vortex detection technology \cite{PRL.87.080402,PRL.81.5477,PRA.60.R3381,PRA.64.031601}. It has also been used to explain the 'zipper' pattern observed in experiment \cite{PRL.95.190403}. These are important topics in studies of the coherence properties of condensates, in developing interferometric phase detection, and in improving measurements of fundamental constants, as well as temporal, gravitational and rotational sensing. The interference of two freely expanding condensate clouds, both containing a vortex, under free expansion is studied in Refs. \cite{PRA.64.031601} and \cite{SSP.108.993}. However, the systematic examination of the role of vortex interactions in the formation of different interference patterns in this process is yet to be carried out.

With the development of atom chip technology \cite{RMP.79.235} an in trap interferometer \cite{PRA.64.063607,PRL.92.050405,PRA.87.023603,PRL.100.100402,PRL.94.090405} is more desirable as the thermal and quantum fluctuations are naturally averaged and the atoms can be recycled. It also provides better contrast than with free expansion. In this paper we present detailed numerical simulations of the in-trap interference of two-dimensional (2D) condensates with vortex states. Since we are interested primarily in the 2D dynamics, an axially symmetric harmonic trap is used in which the radial trapping frequency is much smaller than the axial frequency, to the extent where the axial dimension is sufficiently thin that all dynamics in that direction can be neglected. This highly anisotropic trap can be written in the form of
\begin{eqnarray}
%\begin{split}
V_{trap}\left(\textbf{r}\right) & =& V_{\perp}(x,y)+V_z(z)\\
& =& \frac{1}{2} m \omega_{\perp}^2 \left(x^2 + y^2\right)
+ \frac{1}{2} m \omega_z^2 z^2,
%\end{split}
\end{eqnarray}
with $\omega_z \gg \omega_{\perp}$.

Numerically, a vortex located at $(x_0, y_0)$ with winding number $s$ can be produced by evolving the wave function in imaginary time, subject to the constraints of normalization, and with the phase being given by
\begin{equation}
\theta(x,y)=s\arctan{\frac{y-y_0}{x-x_0}}.
\end{equation}
At zero temperature, the dynamics of a system with $N$ atoms is then governed by the 2D Gross-Pitaeviskii (GP) equation
\begin{equation}\label{GPE-2d}
i\hbar\frac{\partial}{\partial t}\phi=-\frac{\hbar^2}{2m}\nabla_{\perp}^2\phi+V_\perp(x, y) \phi+g_{2D}N\left\vert\phi\right\vert^2\phi,
\end{equation}
where $g_{2D}=2\sqrt{2\pi}a_s a_z\hbar\omega_z$ is the 2D coupling constant, $a_z=\sqrt{\hbar/m\omega_z}$ is the oscillator length in $z$ direction, and $a_s$ is the bulk $s$-wave scattering length. In our simulations of a system of $^{87}Rb$ BECs containing $ N= 2\times10^4$ atoms, we use a reduced $s$-wave scattering length $a_s=0.3\times5.4nm$, to make the core region of the vortex clear to see. The condensates are prepared in a harmonic trap whose axial and radial frequency are $\omega_z=2\pi\times100~{rad\cdot s}^{-1}$ and $\omega_\perp=2\pi\times5~{rad\cdot s}^{-1}$ respectively. At initial time they are displaced away from the trap center for a distance $\Delta x$ and then are released to evolve and form interference patterns. We examine cases of different winding number and initial separation to reveal the vortex interaction in the linear and nonlinear regimes, and the mechanism leading to the formation of specific interference patterns.

The paper is organized as follows. In section 2 we review the situation where one condensate in its ground state collides at the centre of the harmonic trap with another condensate which contains a vortex of charge $s$ at its centre. In section 3 we will discuss the formation of different interference patterns generated by two condensates each containing a singly charged vortex ($s =\pm1$) in the linear and nonlinear regimes. The dynamics of the vortex cores is distinct in different regimes because of the competition between the centre-of-mass (c.m.) kinetic energy of the clouds and the interaction between the vortex cores. A brief summary and discussion is given in the last section.

\section{Collision of condensates with a single vortex state}

\begin{figure}[t]
  \centering
   \includegraphics[scale=0.8, bb=-20 220 776 635]{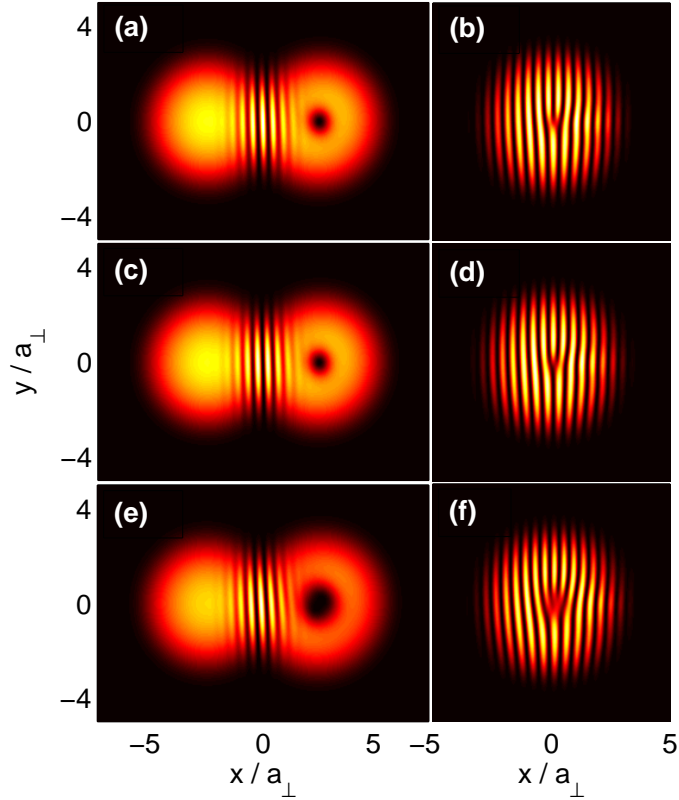} %
   \caption{Density profiles of two condensates with a large initial separation interfering in a harmonic trap. (a,~b) The density profiles of a ground state condensate interfering with another condensate with a singly charged vortex. The initial phase difference at the contact point is $0$. (c,~d) The density profiles of a ground state condensate interfering with another condensate with a singly charged vortex. The initial phase difference at the contact point is $\pi$. (e,~f) The density profiles of a ground state condensate interfering with another condensate with a doubly charged vortex. The initial phase difference at the contact point is $0$. Times are (a,~c,~e) $\tilde t=1.25$, (b,~d,~f) $\tilde t=1.6$. }\label{pg-hv}
\end{figure}

If there is no interaction between atoms, Eq. (\ref{GPE-2d}) turns into a linear Schr\"{o}dinger equation and the system will correspondingly respond linearly. Two noninteracting condensates displaced from the centre of a harmonic trap will form straight and uniform interference fringes. When the interatomic interaction is turned on, the system will either respond linearly or nonlinearly during the dynamical process depending on the competition between the c.m. kinetic energy and the interaction energy of the condensates during the interference. The peak kinetic energy of the condensate is $m\omega_\perp^2\Delta x^2/2$, while under the Thomas-Fermi (TF) approximation the interaction energy of the system is $\mu_{TF}-V_{\perp}(\textbf{r}_\perp)$, provided that the radius of the condensates $r_\perp(x,y)$ is smaller than the TF radius $r_{TF}$, where $r_{TF}=\sqrt{2\mu_{TF}/m\omega_\perp^2}$. At the bottom of the trap $V_\perp$ is zero, which means that if the initial separation of the condensates is comparable to the diameter of each condensate, i.e. $2\Delta x>2r_{TF}$, the peak kinetic energy of the condensates is large and dominates the nonlinearity. In this linear regime the unperturbed condensates always show straight fringes during the interference, as is seen in noninteracting systems.

With the appearance of a vortex state, the initial relative phase $\Delta\theta$ of two clouds is position dependent because of the velocity field with the phase varying over $2s\pi$ around the vortex, which gives the exchange of the bright and dark regions of the interference pattern, as shown in any density plot in Fig. \ref{pg-hv} and Figs. \ref{pp-hv}-\ref{pn-lv}. We find that the interference pattern depends on the initial relative phase of the condensates at the contact point. If the phase difference is 0, the dark region (high density area) is in the middle (see Fig. \ref{pg-hv}(a)). If the phase difference is $\pi$, the bright region (low density area) is in the middle (see Fig. \ref{pg-hv}(c)).

Figure \ref{pg-hv} shows the interference patterns arising from a condensate cloud in the ground state and a condensate cloud in a single vortex state, where the winding number is $s=1$ (Figs. \ref{pg-hv}(a)-\ref{pg-hv}(d)) and $s=2$ (Figs. \ref{pg-hv}(e) and \ref{pg-hv}(f)) respectively. The bright region is a low-density area, while the dark region is a high-density area. The initial displacement for all cases in Fig.\ref{pg-hv} is $\Delta x=7.5a_\perp$, where $a_\perp=\sqrt{\hbar/m\omega_\perp}$. For this initial displacement the maximal c.m. kinetic energy is much larger than the interatomic interaction energy. In Figs. \ref{pg-hv}(a), \ref{pg-hv}(c) and \ref{pg-hv}(e) we can see that the interference fringes at time $\tilde t=t/t_0= 1.25$ , where $t_0=1/\omega_\perp$ is used as the unit of time, look similar to those of two unperturbed condensates with a large initial displacement . However, due to the angular momentum that one of the condensates possesses, the fringes are tilted from the vertical until they reach the position of the vortex core, where the phase slip in fringes appears clearly. When the core region of the vortex enters the interfering region, topological defects in the fringes at the position of the vortex core appear. There are $s$ more fringes on one side of the singularity than on the other side. An observation of this forklike dislocation in the interference fringes is a clear signature of the presence of a vortex \cite{PRL.81.5477}. The interference patterns at $\tilde t = 1.57$ are shown in Figs. \ref{pg-hv}(b), \ref{pg-hv}(d) and \ref{pg-hv}(f).

\begin{figure}[t]
  \centering
   \includegraphics[scale=0.8, bb=-10 310 776 555]{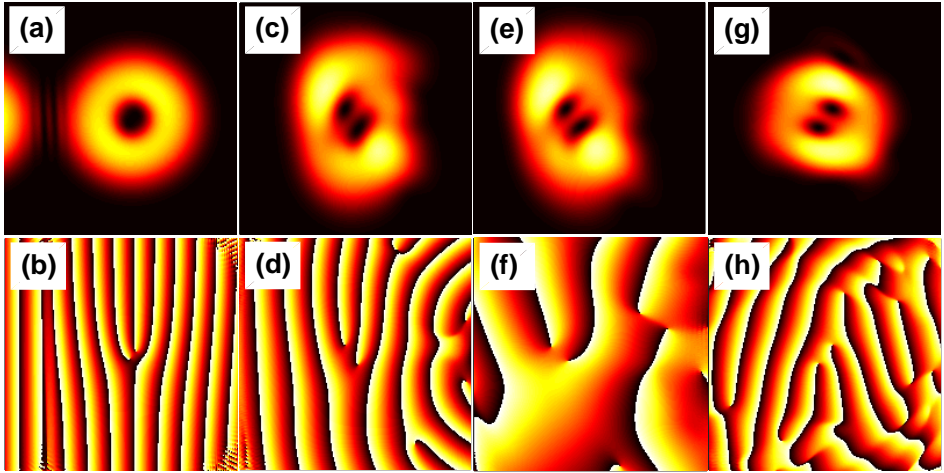} %{fig/vortex2_1.eps}
   \caption{Dissociation of the vortex with charge $s=2$. The top panels are the density profiles of the condensate cloud that originally contain a doubly charged vortex after it has interfered with a ground state condensate. The bottom panels are the corresponding phase diagrams. Times are (a,~b) $\tilde t=1.1$, (c,~d) $\tilde t=2.5$, (e,~f) $\tilde t=3.0$, (g,~h) $\tilde t=3.8$. }\label{vortex2}
\end{figure}

The $s = 2 $ central vortex eigenstate contains a doubly quantized vortex which is unstable under perturbation and will dissociate into two unit vortices ($s=1$) circling around each other in a clockwise direction. The tilted angle of the fringes in Fig.\ref{pg-hv}(e) is more obvious than in Figs. \ref{pg-hv}(a) and \ref{pg-hv}(c). As seen in Fig. \ref{vortex2} density and phase modulations during the collision process make the $s=2$ vortex dissociate into two unit vortices immediately when the two clouds interfere. These vortices remain near the centre of the condensate and circle around each other due to the repulsive force between vortices with the same sign. This is the reason why there is a forklike fringe with three branches on the upper plane in Fig. \ref{pg-hv}(f). The position of the right branch is a bit higher than the left branch due to the fact that the centres of the vortex cores are located at different positions. The dynamical evolution of a doubly quantized vortex imprinted in a BEC has been recently reported in Ref. \cite{PRL.97.180409}.

\section{Collision of condensates with two vortex states}

\subsection{Linear regime}
\begin{figure}[t]
  \centering
   \includegraphics[scale=0.8, bb=-60 240 776 630]{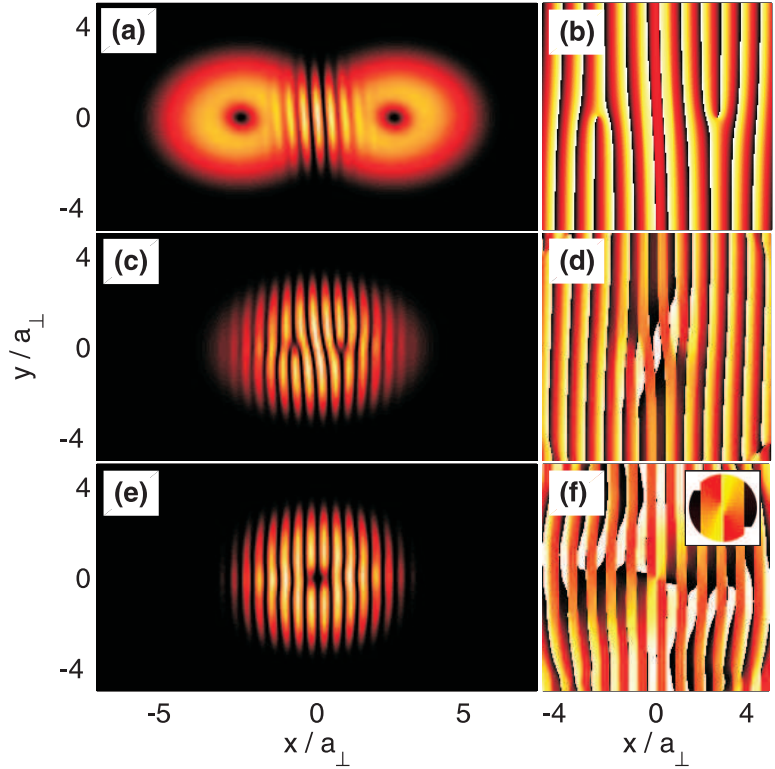} %{fig/pp-hv2_1.eps}
   \caption{Density (left panels) and phase (right panels) profiles of two condensate clouds each containing a vortex with charge $s=1$. The inset in (f) is a zoom plot of the central part of the phase diagram. The initial displacement and relative phase are $\Delta x=7.5a_\perp$ and $\Delta\theta=\pi$ respectively. Times are (a,~b) $\tilde t=1.24$, (c,~d) $\tilde t=1.5$, (e,~f) $\tilde t=1.62$. }\label{pp-hv}
\end{figure}

\begin{figure}[t]
  \centering
   \includegraphics[scale=0.8, bb=-60 240 776 630]{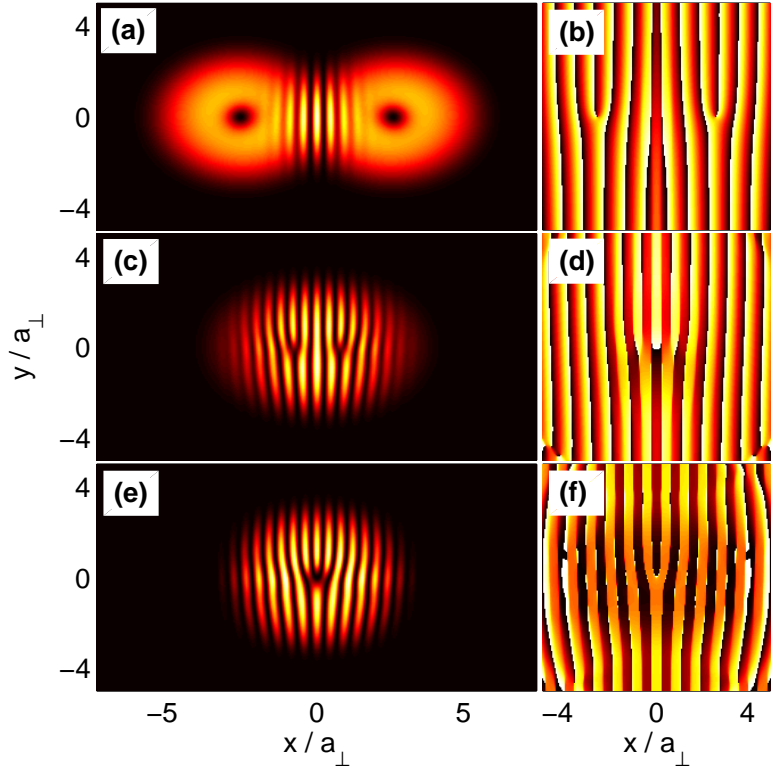} %{fig/pn-hv_1.eps}
   \caption{Density (left panels) and phase (right panels) profiles of two condensate clouds each containing a vortex but with opposite charge. The initial displacement and relative phase are $\Delta x=7.5a_\perp$ and $\Delta\theta=\pi$ respectively. Times are (a,~b) $\tilde t=1.24$, (c,~d) $\tilde t=1.5$, (e,~f) $\tilde t=1.62$.  }\label{pn-hv}
\end{figure}

Now we will consider the situation where there is initially a vortex in each cloud to examine the fringe alternating in the interference process.  We still set the initial displacement $\Delta x=7.5a_\perp$. When the two condensates interfere, the centres of the two vortices are indicated by the two dislocations. If the condensates contain vortices with the same charge, the slope of the fringes is greater than in the situation where there in only one vortex (see Fig. \ref{pp-hv}(a) and Fig. \ref{pg-hv}(a)). The interference pattern is two-fold rotationally symmetric about the centre of the system because the tangential velocity of the two condensates at the contact point is in opposite directions. For the condensates containing two vortices of opposite charge we get straight fringes, as is obtained for the two unperturbed condensates, but the shape of the fringes is not symmetric about the $x$ axis any more because the angular momenta they have are in opposite directions. In Fig. \ref{pn-hv} we can see that the fringes are wider in the $y<0$ half plane. This will been shown more clearly in the case of small separations that follows (see Fig. \ref{pn-lv}), where the spacing of the fringes is large.

In comparison, the peak kinetic energy due to the large displacement is much larger than the interaction energy between the vortex cores, this leads to the two vortices passing straightly through each other no matter the charges they carried. This is clearly shown in Figs. \ref{pp-hv}(e), \ref{pp-hv}(f), \ref{pn-hv}(e) and Fig. \ref{pn-hv}(f), especially in the phase diagrams, where the points with phase singularities indicate the positions of the centres of the vortex cores. The vortices move towards each other with the centres of their cores fixed on the $x$ axis. As shown in Fig. \ref{pp-hv} and Fig. \ref{pn-hv}, the core region of the two vortices overlaps as the condensates reach their maximal overlap. In Fig. \ref{pp-hv}(f), we can see that the two singularities are nearly at the same point, acting the same as a doubly charged vortex. In Fig. \ref{pn-hv}(f), the two singularities have vanished which means that the two vortices of opposite charge have annihilated, and a curved soliton forms instead. In both cases the cores of the vortices will separate and move apart from each other thereafter.

We note that the forklike fringe in Fig. \ref{pn-hv}(e) looks like that in Fig. \ref{pg-hv}(f), but is intrinsically different because of the formation mechanism. The former is formed by the creation of a curved soliton, leading to the complete separation between the branch in the middle and the others by a deep depression of density. The interference pattern is still symmetric about the $y$ axis. The latter is formed due to the two vortices located at different places, which makes the three branches connect, and the pattern is not symmetric about the $y$ axis.

\begin{figure}%[H]
  \centering
   \includegraphics[scale=0.8, bb=-60 200 776 660]{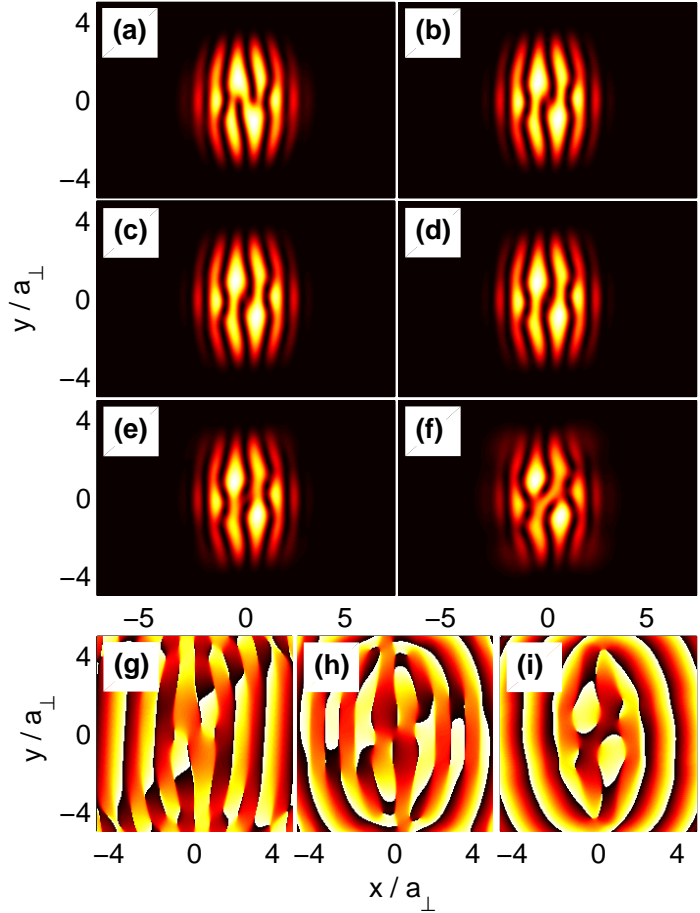} %{fig/pp-lv_1.eps}
   \caption{Density profile (a)-(f) and phase diagram (g)-(i) of two condensates each containing a vortex with same charge $s=1$. The condensates start with a small initial separation interfere and collide in a harmonic trap. The initial displacement and relative phase are $\Delta x=4.0a_\perp$ and $\Delta\theta=0$ respectively. Times are $\tilde t=1.52,~1.6,~1.62,~1.64,~1.7,~1.8$ from (a) to (f). (g)(h)(i) are the corresponding phase diagrams of (a)(d)(f).}\label{pp-lv}
\end{figure}

\subsection{Nonlinear regime}
\begin{figure}[t]
  \centering
   \includegraphics[scale=0.8, bb=-60 200 776 660]{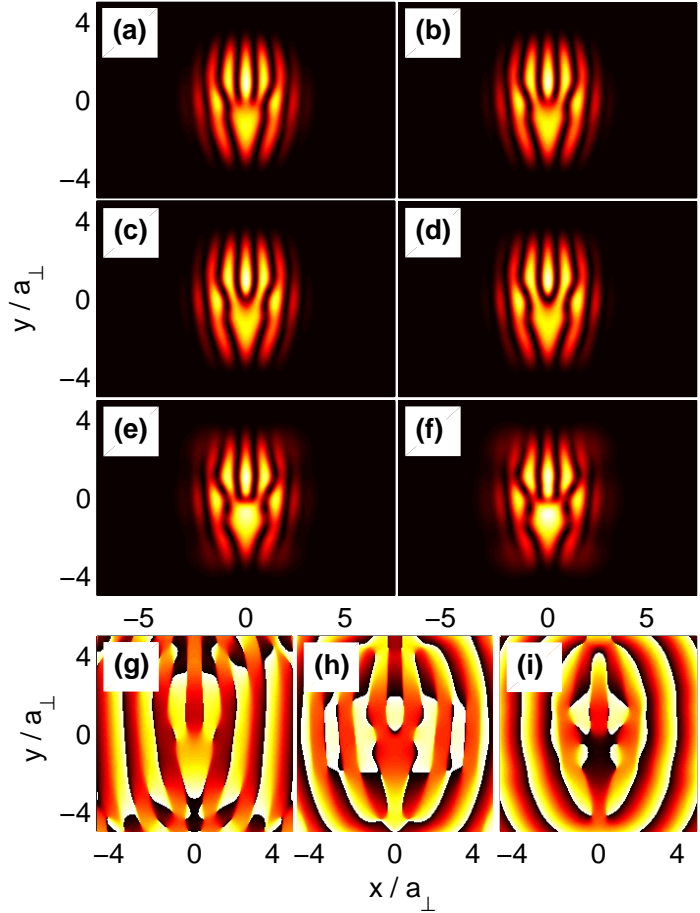} %{fig/pn-lv_1.eps}
   \caption{Density profiles (a)-(f) and phase diagram (g)-(i) of two condensates each containing a vortex but with opposite charge. The initial displacement and relative phase are $\Delta x=4.0a_\perp$ and $\Delta\theta=0$ respectively. Times are $\tilde t=1.52,~1.56,~1.6,~1.62,~1.78,~1.82$ from (a) to (f). (g)(h)(i) are the corresponding phase diagrams of (a)(d)(f).}\label{pn-lv}
\end{figure}

As the initial displacement of the condensates decreases ($\Delta x = 4a_\perp$), the peak kinetic energy of the condensates decreases accordingly. When the interaction energy of the condensates is comparable to the peak c.m. kinetic energy, we reach the nonlinear regime. The central interference fringes are wider and the curvature of the fringes is larger than that of the systems we discussed above (see Figs. \ref{pp-lv} and \ref{pn-lv}). Compared with the inference of two ground state clouds starting with same displacement as in Ref. \cite{PRA.87.023603}, the expansion of the condensates is much more obvious, implying that the scattering is enhanced by the presence of angular momentum. In the presence of a vortex, the system is more sensitive to the nonlinearity. After the first interference, the fringes have been strongly distorted by excitations.

When the two condensates both contain a vortex, the motion of the vortex cores during the interference is of particular interest. When the two vortices have the same charge, they repel each other as they approach. In Fig. \ref{pp-lv}, we present a sequence showing the continued evolution of the vortices during the interference process of two condensates. The two vortices make an "H" shape pattern which has been observed in experiment \cite{PRA.64.031601}. The centres of the vortex cores in the two condensates deviate slightly from the $x$ axis when they move towards each other. This is different from the linear case where the cores of the vortices are always on the $x$ axis. The kinetic energy of the cores is not large enough to overcome the repulsive potential between the vortices to make them overlap and pass through each other.

In the situation where the two vortices have opposite charge, the dynamics of the cores of the vortices are the same as in the linear regime until the point where more vortices are excited by the nonlinearity. The time evolution is given in Fig. \ref{pn-lv}.  The cores of the two vortices move towards each other at the same level, i.e. along the $x$ axis, (see Fig. \ref{pn-lv}(a)-(c)) and overlap (see Fig.\ref{pn-lv}(d)), forming a bent soliton. In Figs. \ref{pn-lv}(g)-\ref{pn-lv}(i), we give the phase diagrams at $\tilde t=1.52,~1.64,~1.8$, which correspond to the times before, at and after the cores of the two vortices overlap.

\section{Conclusion}

In conclusion, we have investigated the role of different winding number and initial displacement of well separated 2D trapped BECs during an interference process, with special attention paid to the situation with two vorticity states. There are two distinct dynamical regimes given by the competition between the interaction energy and the kinetic energy of the condensates; When the peak kinetic energy, which is decided by the initial displacement of the condensates, is much larger than the interatomic interaction energy, the interference is linear, and the interference pattern is similar as that of the noninteracting case. The two vortices can pass through each other with the centre of the core regions being fixed on the $x$ axis. The core of the two vortices with the same winding number can overlap and briefly form a vortex with a higher winding number before separating. When the initial displacement of the condensates decreases, the peak kinetic energy decreases accordingly leading to the interaction term being important as it is comparable to the peak kinetic energy. The nonlinearity changes the dynamics of the system. In this nonlinear regime, the cores of the two vortices of the same winding number can not overlap due to the repulsive force between them. they pass by each other and make the interference pattern a ``H'' shape. For two vortices of opposite winding number they will annihilate, forming a soliton which makes a fork like interference pattern accordingly.

\section*{Acknowledgements}
  We would like to thank Derek Lee and Andrew Armour for many useful discussions. We acknowledge support through the EPSRC. T.Y. also acknowledges support through NSFC11347025, NSFC11347605 and the Science Foundation of Northwest University (No.13NW16).
\section*{References}
%\bibliographystyle{unsrt}
%\bibliography{ThesisEx}
%\bibliography{Vortex_Interactions.bbl}

\end{document}